\documentclass[12pt]{article}
\usepackage{graphicx}
\usepackage{amsmath}
\usepackage{amssymb}

\newcommand\pubnumber{NuPhys2017-Roskovec}
\newcommand\pubdate{\today}

\def\napoli{Pontificia Universidad Cat\'{o}lica de Chile \\
Instituto de F\'{i}sica, Avda. Vicu\~{n}a Mackenna 4860, Santiago, CHILE}

\def\Title#1{\begin{center} {\Large #1 } \end{center}}
\def\Author#1{\begin{center}{ \sc #1} \end{center}}
\def\Address#1{\begin{center}{ \it #1} \end{center}}

\newcommand\pubblock{\rightline{\begin{tabular}{l} \pubnumber\\
         \pubdate  \end{tabular}}}
\newenvironment{Abstract}{\begin{quotation}  }{\end{quotation}}
\newenvironment{Presented}{\begin{quotation} \begin{center} 
             PRESENTED AT\end{center}\bigskip 
      \begin{center}\begin{large}}{\end{large}\end{center} \end{quotation}}

\def\beq{\begin{equation}}
\def\eeq#1{\label{#1}\end{equation}}
\def\eeqn{\end{equation}}

\def\beqa{\begin{eqnarray}}
\def\eeqa#1{\label{#1}\end{eqnarray}}
\def\eeqan{\end{eqnarray}}

\let\bar=\overbar

\def\Dslash{\not{\hbox{\kern-4pt $D$}}}
\def\dslash{\not{\hbox{\kern-2pt $\del$}}}

\def\msb{{\bar{\ssstyle M \kern -1pt S}}}

\begin{document}
\begin{titlepage}
\pubblock

\vfill
\Title{Reactor Neutrino Spectrum and Flux Measurement}
\vfill
\Author{ Bed\v{r}ich Roskovec} 
\Address{\napoli}
\vfill
\begin{Abstract}
Reactor neutrinos play a substantial role in the study of the fundamental properties of neutrinos. With current and upcoming precision experiments, it is essential more than ever to understand the reactor neutrino flux and spectrum. However, two discrepancies between prediction and measurement are observed. On the one hand, there is a $\sim$6\% total measured flux deficit, known as the reactor antineutrino anomaly, consistently seen by several experiments at short baselines. On the other hand, there is an observed excess over prediction for $\bar\nu_e$ energies between $5-7\text{ MeV}$. We discuss current status of the experimental measurements and provide an outlook.
\end{Abstract}
\vfill
\begin{Presented}
NuPhys2017: Prospects in Neutrino Physics\\
Barbican Centre, London, UK,  December 20--22, 2017
\end{Presented}
\vfill
\end{titlepage}
\def\thefootnote{\fnsymbol{footnote}}
\setcounter{footnote}{0}

\section{Introduction}
Nuclear reactors are a powerful source of pure low energy electron antineutrinos. They have been used to study fundamental neutrino properties over decades spanning from the discovery of neutrino, precisely electron antineutrino, in Savannah River experiment \cite{reines1}, through a period of measurements in the 80s and 90s \cite{80s1, 80s2, 80s3, 80s4, 80s5} with a series of short baseline experiments, until the first observation of $\Delta m^2_{21}$-driven neutrino oscillations measured by KamLAND experiment \cite{Kamland}.

The current generation of large scale reactor neutrino experiments (Daya Bay, RENO, Double Chooz) ware designed to study $\bar\nu_e$ disappearance at distances $\sim$1~km and discovered new mode of neutrino oscillations expressed by a non-zero value of the $\theta_{13}$ mixing angle \cite{DYB_first, RENO_first, DC_first}. These experiments use near and far detectors to cancel out systematics from the reactor neutrino prediction. The unprecedentedly large statistics of antineutrinos collected in these near detectors, as an example Daya Bay's more then 2.2 million events  \cite{DYB_long}, allows to study in detail the reactor neutrino flux and spectrum. The experiments have confirmed the so-called reactor antineutrino anomaly, when they measured about 6\% lower overall flux compared to the reevaluated prediction from 2011 \cite{Huber, Mueller}. Daya Bay recently provided a closer look to the measured deficit, being able to disentangle contribution from particular isotopes while investigating nuclear fuel evolution. In addition to the flux discrepancy, these experiments, as well as other short baseline experiments such as NEOS \cite{NEOS_bump}, observed statically significant excess over prediction in the $\bar\nu_e$ spectrum at the energy range of $5-7\text{ MeV}$. This `bump' was linked to the reactor neutrino production process.

In this note, we give a brief introduction to reactor neutrinos, describe the basics of two prediction methods and $\bar\nu_e$ detection. We then focus on the observed anomalies and mention possible explanations. We conclude with the outlook to upcoming experiments, which are designed to shed more light on those anomalies.

\section{Reactor Neutrinos}
Nuclear reactors use fission of heavy nuclei to produce thermal energy. There are two general types of reactors that differ by their fuel content. Research reactors typically use highly enriched uranium (HEU), where almost exclusively only $^{235}$U is burned. On the other hand, commercial reactors typically use low enriched uranium (LEU) where the situation is more complex. There are four main isotopes whose fissions account for 99.9\% nuclear reactor power, namely $^{235}$U, $^{238}$U, $^{239}$Pu and $^{241}$Pu. In addition, the fuel composition is changes significantly during the fuel cycle, which is typically a few months long. While in the beginning, mostly $^{235}$U is burned, the plutonium isotopes are building up and at the end of the fuel cycle $^{239}$Pu is the largest contributor to the total $\bar\nu_e$ flux.

The fission products of four main isotopes are neutron rich isotopes, which undergo a series of beta decays to reach stability. Beta decay can be schematically expressed as $^A_NX\rightarrow^A_{N-1}Y+e^-+\bar\nu_e$. In this processes, pure electron antineutrinos are emitted. There are $\sim$$6\:\bar\nu_e\text{'s}$ produced per fission, which is about $2\times10^{20}\:\bar\nu_e/\text{s/GW}_{th}$. This makes nuclear reactors the most powerful man-made source of electron antineutrinos.

There are two complementary methods to predict reactor neutrino spectrum. The idea behind the 'ab initio' summation method it to sum over all decay branches of all possible fission isotopes with appropriate weights to obtain the aggregate antineutrino spectrum. The data are taken from databases. This method however has to deal with large uncertainties for some important fission products. In addition, not all branching ratios are known. This results in a larger uncertainty for the summation method compared to the second one, the conversion method. Conversion method uses the electron spectra for $^{235}$U, $^{239}$Pu and $^{241}$Pu measured at ILL \cite{ILL1, ILL2, ILL3} and convert them into antineutrino energy. Later the electron spectrum for $^{238}$U was measured as well \cite{U238}. The electron spectra for particular isotopes are intrinsically already summed over all decays. They are then fitted by number of virtual branches. One has to make assumptions about the shape of these branches, taking into account several aspects such as forbidden decays, etc. It was actually the reevaulated conversion method for $^{235}$U, $^{239}$Pu and $^{241}$Pu isotopes \cite{Huber} with the calculation for $^{238}$U \cite{Mueller}, commonly referred to as Huber+Mueller model, which gave rise to the reactor antineutrino anomaly. The uncertainty for the conversion method is smaller and therefore it is the leading method used nowadays. 

The primary reaction used for antineutrino detection used in ongoing experiments is inverse beta decay (IBD) $\bar\nu_e+p\rightarrow e^++n$. This reaction has a threshold at 1.806~MeV. It is used due to its relatively large cross-section and coincidence of the prompt signal formed by the positron and delayed neutron capture. This distinct signature is powerful to suppress backgrounds. Moreover, the prompt energy is directly linked to the initial antineutrino energy though $E_{\bar\nu_e}\simeq E_{prompt}+(1.806-2\times0.511\text{ MeV})$, where the threshold and positron annihilation is reflected. This relation is useful when reading the experimental results since those are usually shown using prompt signal energy.

\section{Reactor Antineutrino Anomaly}
\begin{figure}[htb]
\centering
\includegraphics[height=2.5in]{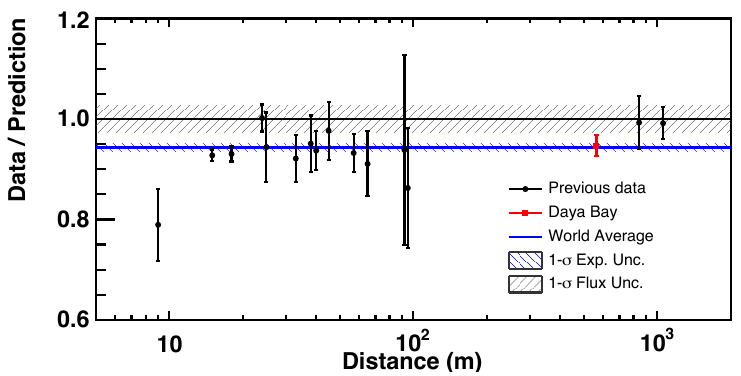}
\caption{The ratio of observation over the Huber+Mueller model prediction of past measurements corrected for the current knowledge of three-neutrino oscillations. The Daya Bay measurement of $R=0.946\pm0.020\text{ (experimental)}$ \cite{DYB_spectrum} is highlighted. The global average $R=0.943\pm0.008\text{ (experimental)}\pm0.023\text{ (model)}$ \cite{DYB_spectrum} indicates a deficit in the measured flux. The figure does not include recent RENO result which can be found in \cite{RENO_react}.}.
\label{fig:react_anomaly}
\end{figure}

A number of experiments measured reactor neutrino flux at distances $\mathcal{O}(10-100\text{ m})$. While the results are consistent among experiments, the comparison with the Huber+Mueller model revealed a deficit of about 6\%. The situation is graphically illustrated in Figure~\ref{fig:react_anomaly}, where the ratios of past measurements over prediction corrected for currently known three-neutrino oscillations are shown. With the Daya Bay result, a world average ratio $R=0.943\pm0.008\text{ (experimental)}\pm0.023\text{ (model)}$ \cite{DYB_spectrum} is obtained. 

This deficit in the measured flux can be in principle explained by the existence of one sterile neutrino additional to the common three active neutrino flavors. Sterile neutrinos do not participate in the weak interactions but still could take part in neutrino oscillations. The existence of sterile neutrino, which can explain the reactor anomaly, would imply the existence of additional mass state with the mass squared difference $\Delta m^2_{new}\gtrsim1\text{ eV}^2$. Such a value would lead to the fast neutrino oscillations with oscillation length $\sim$$\mathcal{O},(\text{m})$, whose signature would average out in current and past detectors, due to the baseline and energy resolution. The averaging effect manifests itself as mere flux deficit when a fraction of the $\bar\nu_e$'s changes flavor to sterile state. More credibility to the existence of sterile neutrinos is added by the observation of other discrepancies at several experiments: LSND \cite{LSND}, MiniBooNE \cite{MiniBooNE}, SAGE \cite{SAGE} and GALLEX \cite{GALLEX}. Those can be as well plausibly explained by sterile neutrinos.

The conservative explanation of the reactor anomaly is an underestimations in the prediction. To resolve whether there are sterile neutrinos, it is essential to observe typical $\frac{L}{E}$ dependence pattern. New precise experiments located close to the antineutrino source are needed. 

\section{Excess in the Reactor Neutrino Spectrum}
\begin{figure}[htb]
\begin{center}
\includegraphics[width=0.53\textwidth]{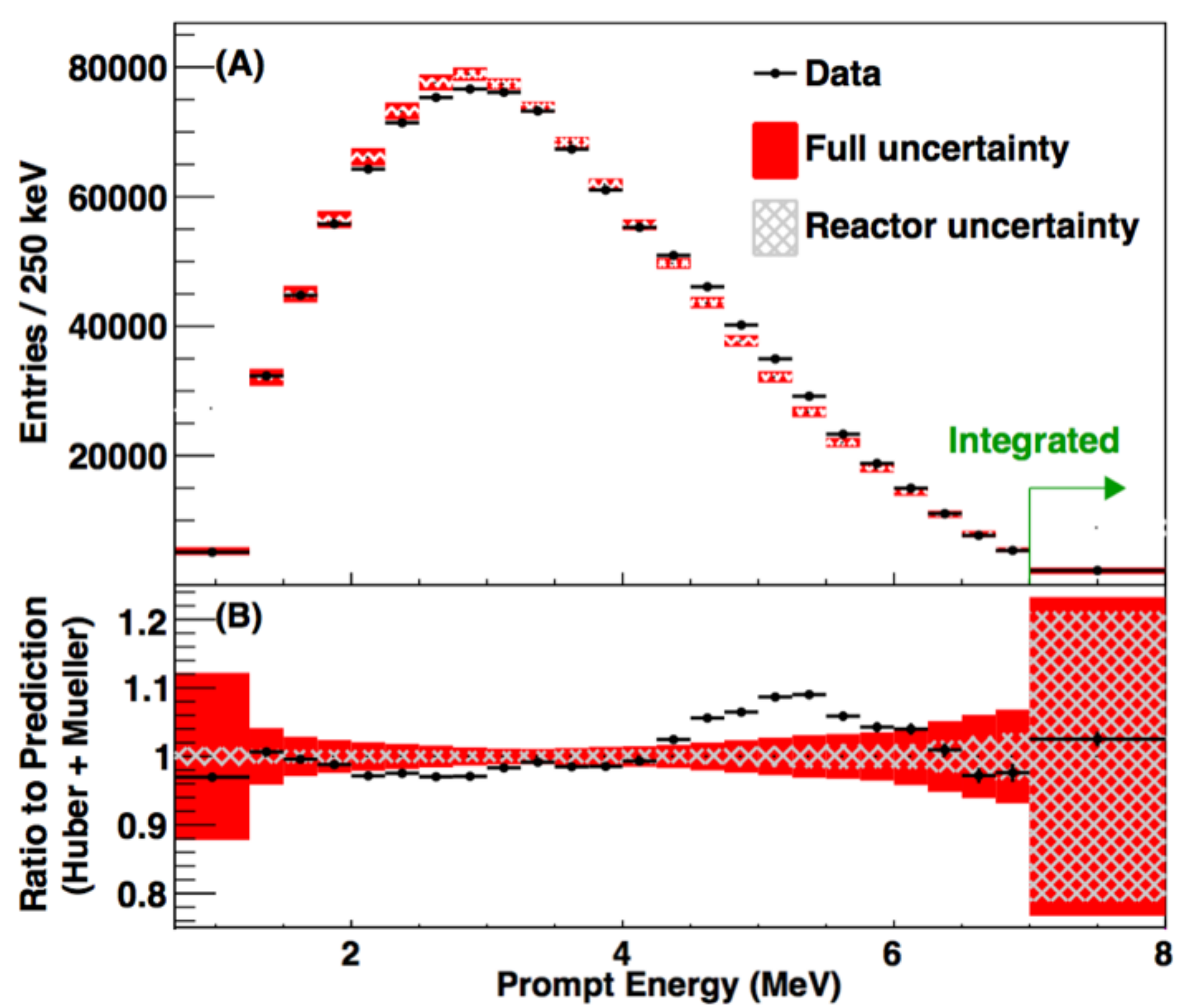}
\includegraphics[width=0.42\textwidth]{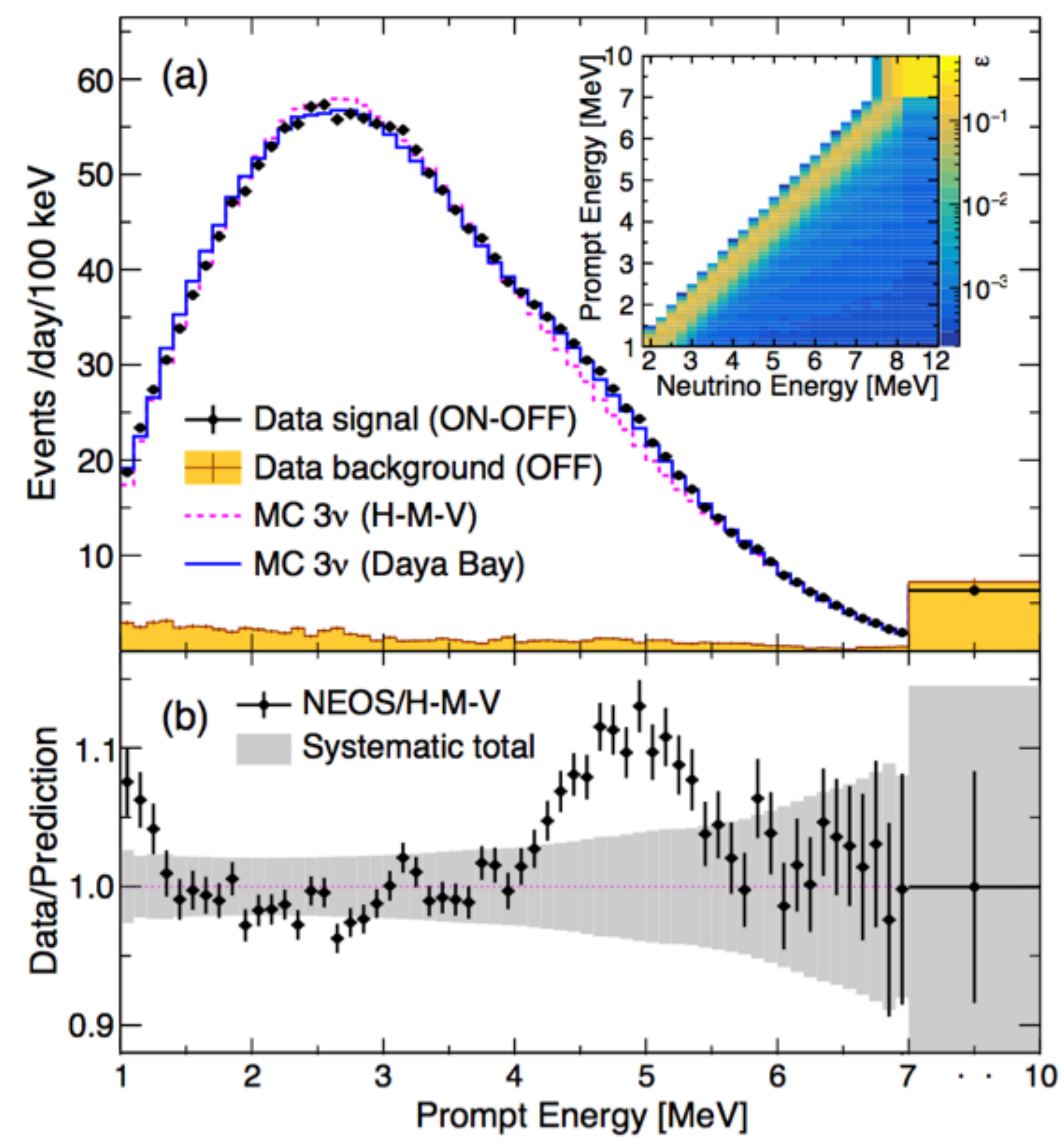}
\caption{\label{fig:bump} Examples of the excess of measured reactor neutrino spectrum over prediction observed by the Daya Bay experiment \cite{DYB_spectrum} (left) and by the NEOS experiment \cite{NEOS_bump} (right).}
\end{center}
\end{figure}

The large statistics of current experiments allow testing the reactor neutrino energy spectrum shape with unprecedented precision. The comparison with the prediction exhibits an excess of detected antineutrinos in the energy range of $5-7\text{ MeV}$ ($4-6\text{ MeV}$ IBD prompt energy). This feature was observed by several experiments \cite{DYB_spectrum, RENO_bump, NEOS_bump, DC_bump} and an example is shown on Figure~\ref{fig:bump}. The most precise measurement comes from Daya Bay, and has local significance of the `bump' at the level of 4.4$\sigma$~\cite{DYB_spectrum}. This local excess cannot be explained by the existence of sterile neutrinos since those would alter the spectrum for the whole energy range. As well as bump is not a systematic bias related to the detector since it was not observed in other types of events such as spallation $^{12}$B spectrum \cite{DYB_spectrum}. On the other hand, it was demonstrated that the size of the excess is correlated with the reactor power \cite{RENO_bump} indicates that the culprit must be sought in antineutrino production. Due to the fact that the bump was not observed in any of the electron spectra used for the conversion \cite{ILL1, ILL2, ILL3}, two implications can be made: either the bump is coming from $^{238}$U, which is more loosely constrained than the other isotopes, or it can come from any isotope if previous measurements were not accurate enough. There might be a difference in the hardness of neutrons used in the measurements with respect to that present in nuclear reactors. The impact of different neutron energy on the final spectrum is still unclear. Either way, further measurements are needed to identify the cause of the bump.

\section{Closer Look to Reactor Antineutrino Anomaly with Fuel Evolution at Daya Bay}
\begin{figure}[htb]
\centering
\includegraphics[width=0.49\textwidth]{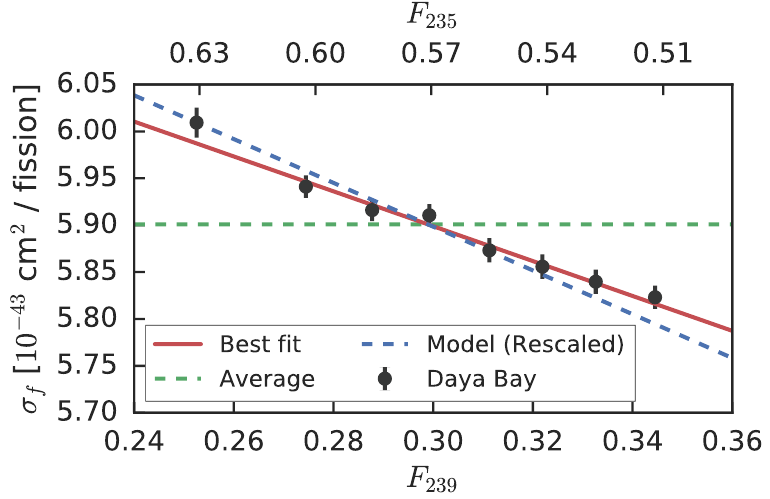}
\includegraphics[width=0.49\textwidth]{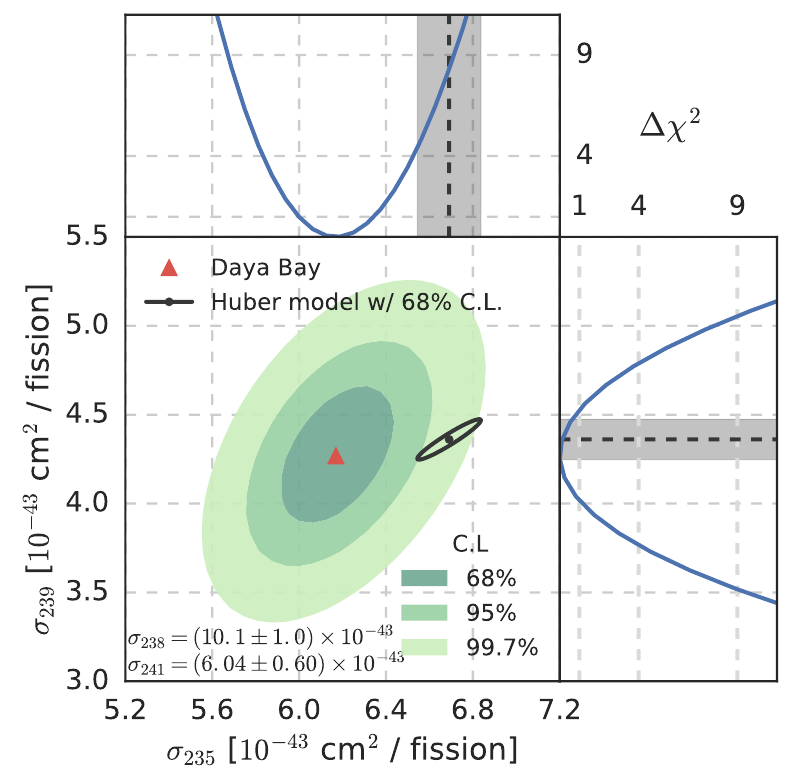}
\caption{The evolution of reactor neutrino flux expressed as IBD yield per fission as a function of effective $^{239}$Pu fission fraction \cite{DYB_evolution} (left). The slope of Daya Bay measurement (red line) does not match the predicted evolution of Huber+Mueller model (blue curve), which was corrected for the reactor antineutrino anomaly effect. The Daya Bay measured IBD yield per fission of $^{235}$U and  $^{239}$Pu \cite{DYB_evolution} and the comparison with Huber+Mueller prediction model \cite{Huber, Mueller} (right). The measurement is clearly lower than the prediction for $^{235}$U while $^{239}$Pu is in good agreement.}
\label{fig:evo}
\end{figure}

The relative contribution of four main isotopes changes during fuel cycle in the LEU nuclear reactors. While in the beginning, most of the fissions come from $^{235}$U, at the end built-up $^{239}$Pu takes over this role. The other two isotopes have more or less a constant contribution throughout the cycle. Since the IBD yield per fission is not same for all the isotopes, we can expect change in the overall antineutrino flux due to change of fuel content. Recently, Daya Bay reported a measurement of reactor neutrino flux and spectrum with fuel evolution with unprecedented precision \cite{DYB_evolution}. There is a tension between the measurement and evolution predicted Huber+Mueller model.  Although the spectrum shape evolution agrees within current experimental precision, the evolution of total antineutrino flux does not follow the prediction, see left panel of Figure~\ref{fig:evo}. Furthermore, unique analysis of changing antineutrino flux allowed the experiment to disentangle the contribution of the single isotopes. Daya Bay found that there is a significant deficit between the predicted and measured IBD yield per fission for $^{235}$U while $^{239}$Pu agrees very well  with the model as shown on right panel of Figure~\ref{fig:evo}. If sterile neutrinos are responsible for reactor anomaly, the same deficit should be observed in IBD yield for all isotopes. Since $^{235}$U has significantly lower deficit then $^{239}$Pu, the sterile neutrino hypotheses is weakened. The equal-deficit hypothesis as an explanation for reactor anomaly is disfavoured by Daya Bay on 2.8$\sigma$. However, Daya Bay did not rule out sterile neutrinos completely. For example \cite{Hayes, Bryce_paper} shown that combination of Daya Bay and global data prefers composite model of revisited prediction for $^{235}$U and the existence of sterile neutrinos. Further investigation with current and upcoming experiment is needed to provide a final solution of the reactor antineutrino anomaly.

\section{Outlook}
\begin{figure}[htb]
\centering
\includegraphics[width=0.6\textwidth]{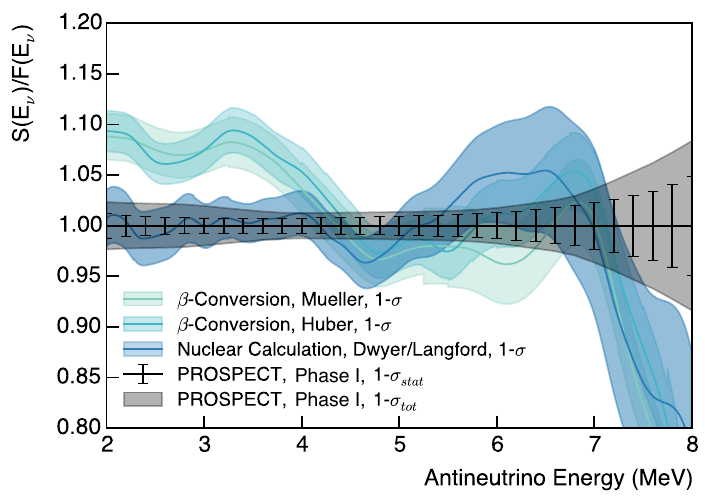}
\includegraphics[width=0.39\textwidth]{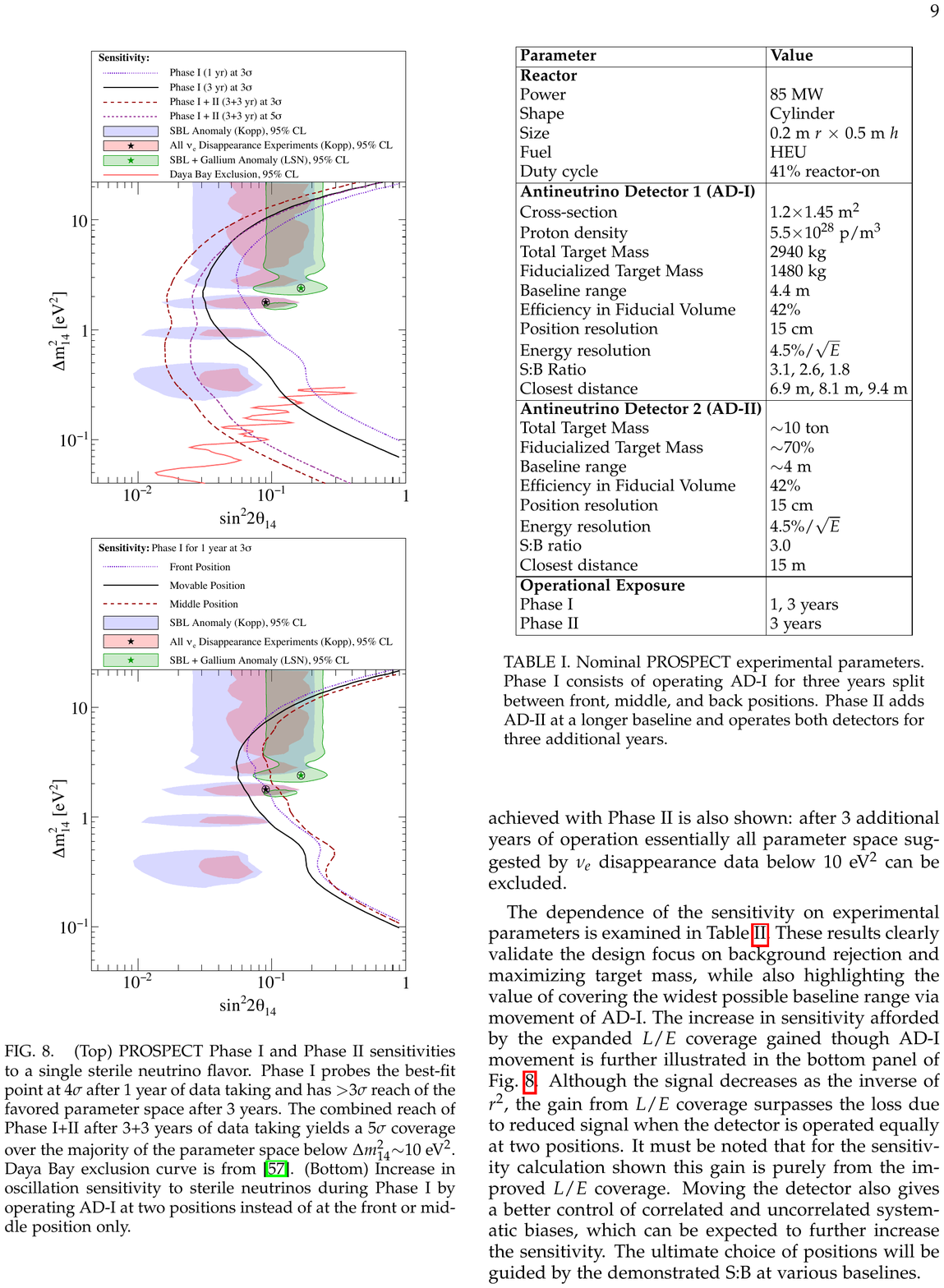}
\caption{The expected sensitivity of the PROSPECT \cite{PROSPECT} to the shape of $^{235}$U spectrum and comparison with several prediction models (left). The expected sensitivity of the experiment to the sterile neutrino mixing \cite{PROSPECT} (right).}
\label{fig:prospect}
\end{figure}

Ongoing experiments will keep improving their precision. For example, the Daya Bay experiment is due to run until 2020, essentially doubling the statistics of the latest analyses \cite{DYB_long, DYB_evolution} and decreasing systematic uncertainties. Nevertheless, new short baseline neutrino experiments are needed to scrutinize the observed anomalies. Several are already taking data, e.g. DANSS \cite{DANSS}, PROSPECT \cite{PROSPECT}, STEREO \cite{STEREO} etc. Without loss of generality we mention the PROSPECT experiment, which will undoubtedly play crucial role in the improvement of our current knowledge. PROSPECT uses research HEU reactor. The experiment thus can directly test the $^{235}$U flux and test the prediction as well as Daya Bay result \cite{DYB_spectrum}. Moreover, the spectrum of $^{235}$U can be studied with aim to search for the bump. The expected sensitivity of the PROSPECT to the shape of $^{235}$U spectrum and its comparison with several predictions is shown in the left panel of Figure~\ref{fig:prospect}. We have already mentioned that there is a possibility that the bump could come from any isotope if the electron spectra measurements at ILL suffer from having used different neutron energies. If the bump is not observed in PROSPECT, we can rule out $^{235}$U and other isotopes are likely its source. 

Also, PROSPECT can with its segmented detector located at the distance of $7-12\text{ m}$ from the reactor search for an $\frac{L}{E}$ oscillation pattern, which would be an unquestionable signal of sterile neutrinos. It can directly test the best fits for reactor anomaly and the anomaly observed by SAGE and GALLEX, see right panel of Figure~\ref{fig:prospect}. Combining with other experiments similar to what was done in \cite{DYB_combination} it can further improve test of the sterile neutrino claims done LSND and MiniBooNE. 

\section{Conclusion}
Reactor antineutrinos were, are and will be great tool for investigating neutrino properties. The era of precise measurements has revealed several deviations from our prediction models, referred to as anomalies. There is a measured deficit in the overall flux as well as in the particular flux of $^{235}$U and there is excess in the reactor antineutrino spectrum in the energy range of $5-7\text{ MeV}$. Current and upcoming experiments will provide more information in the near future. The ultimate resolution of these anomalies will either result in the discovery of new physics or in a revision of our prediction models. In both cases a bright future lies ahead full of new interesting results.

\end{document}